\documentclass[12pt]{article}
\usepackage{epsf}
\usepackage{cite}
\def\beq{\begin{eqnarray}}
\def\eeq{\end{eqnarray}}

\begin{document}
\begin{flushright}
NYU-TH/01/05/15 \\
\end{flushright}

\vspace{0.1in}
\begin{center}
\bigskip\bigskip

{\large \bf A new mechanism for generating density perturbations from inflation}

\vspace{0.4in}      

Gia Dvali$^a$, Andrei Gruzinov$^{a}$, and Matias Zaldarriaga$^{a,b,c}$
\vspace{0.1in}

$^a$ {\baselineskip=14pt \it 
Center for Cosmology and Particle Physics\\[1mm]
}{\baselineskip=14pt \it 
Department of Physics, New York University,  
New York, NY 10003\\[1mm]
}
\vspace{0.1in}
$^b$ {\baselineskip=14pt \it 
 Department of Physics \\[1mm]
}{\baselineskip=14pt \it Harvard University,  
Cambridge, MA 02138\\[1mm]
}
\vspace{0.1in}
$^c$ {\baselineskip=14pt \it 
Department of Astronomy  \\[1mm]
}{\baselineskip=14pt \it Harvard University,  
Cambridge, MA 02138\\[1mm]
}
\vspace{0.1in}
\end{center}

\begin{center}
{\bf Abstract}
\end{center}
\vspace{0.3cm}
We propose a new mechanism to generate density perturbations in inflationary models. Spatial fluctuations in the decay rate of the inflaton field to ordinary matter lead to fluctuations in the reheating temperature. We argue that in most realistic models of inflation the coupling of the inflaton to normal matter is determined by the vacuum expectation values of fields in the theory. If those fields are light during inflation (this is a generic situation in the minimal models of supersymmetric inflation) they will fluctuate leading to density perturbations through the proposed mechanism. We show that these fluctuations could easily dominate over the ones generated through the standard mechanism. The new scenario has several consequences for inflation model building and observations. The proposed mechanism allows to generate the observed level of density perturbations with a much lower scale of inflation and thus generically predicts a smaller level of gravitational waves. The relation between the slope of the spectrum of the produced density perturbations and the potential of the inflaton field is different from the standard relations obtained in the context of slow roll inflation. Because the field responsible for the fluctuations is not the inflaton, it can have significantly larger self couplings and thus density perturbations could be non-Gaussian.  The non-Gaussianity can be large enough to be detectable  by CMB and Large Scale Structure observations. 

\newpage

\section{Basic Mechanism}

In the standard picture \cite{review},
the 
observed density perturbations are produced as follows. As the inflaton field $\phi$ rolls down its potential it is effectively massless, so it fluctuates up and down. These fluctuations are different in different regions, resulting effectively in inflation lasting  different amounts of time in different places. 
 At the end of inflation, the inflaton oscillates 
about the minimum of its potential and decays, reheating the universe. As a result of the fluctuations each region of the universe goes through the same history but at slightly different times leading to adiabatic density perturbations. 

Our idea is different.  To reheat the universe the inflaton has to 
couple to the ordinary particles through a coupling schematically given by,
\begin{equation}
 \lambda\, \phi\, q\, q\,
\end{equation}
where $q$ stands for the ordinary particles (e.g. quarks, leptons and their scalar superpartners),
and $\lambda$ is the coupling strength, which is assumed to be
{\it constant} in the standard picture.  
For simplicity we assume that the density perturbations produced
at the inflationary stage are negligible. So inflation ends at the same time everywhere in the universe. The density fluctuations are created during the reheating
process, because of  fluctuations in the coupling ``constant''
$\lambda$. Since $\lambda$ controls the efficiency by which the
energy stored in the $\phi$-condensate gets converted into radiation,
the fluctuations in $\lambda$ translate into the fluctuations in the final reheating
temperature. Note that in our scenario different regions go through slightly different histories, as the couplings are different. Thermal equilibrium ensures that we get adiabatic fluctuations. 

The key point of our approach is to notice that  $\lambda$ is not a constant,
and can fluctuate in space. The reason why $\lambda$ could fluctuate is that in
supersymmetric theories as well as in the theories inspired
by superstrings,
the effective couplings are not constants, but rather  functions
of the scalar fields in the theory. In the early universe, these scalar fields can
 assume different values in different regions. We shall come back to this issue below.

 Assuming that $\lambda$ is a stochastic variable, the new mechanism works
as follows. The decay rate of the inflaton field is
\begin{equation}
 \Gamma \sim \lambda^2\, m
\end{equation}
where $m$ is the inflaton mass at its minimum. We shall assume that 
$\Gamma$ is less than the Hubble parameter during inflation. If this were not the case and $\Gamma >> H$ then the reaheating would be instantaneous and all the energy stored in the inflaton would immediately be converted into radiation. In that case the reheating temperature would be independent of $\Gamma$. 

When $\Gamma < H$ during inflation, the reheating
temperature is roughly
\begin{equation}
\label{tr}
 T_R \sim \,  \sqrt{\Gamma M_{PL}} \sim \, \lambda \sqrt{mM_{PL}}.
\end{equation}
If $\lambda$ fluctuates, the corresponding fluctuations in the reheating
temperature are:
\begin{equation}
 {\delta\, T_R \over T_R}  \sim {\delta\, \Gamma \over \Gamma} \sim {\delta\, \lambda \over \lambda}
\end{equation}
The fluctuations are completely determined by the fluctuations of $\Gamma$ 
not of $\phi$.

In order to understand  in more detail 
how our mechanism works we will consider a toy model. We write down the full set of perturbation equations in a later section. Let us follow 
the cosmological evolution in two different domains of the Universe in which
$\lambda$ takes two different values equal to $\lambda_1$ and 
$\lambda_2$ respectively. For definiteness we shall assume that
$\lambda_1 > \lambda_2$. According to our assumption,
right after inflation the energy densities in the two domains are equal,
and evolve as non-relativistic matter,
\begin{equation}
 \epsilon_{1matter}\, = \, \epsilon_{2matter} \, \sim \, {1 \over a^3}
\end{equation}
where $a$ is the scale factor. The inflaton decay in each domain
takes place when
\begin{equation}
 \Gamma \,  = \, H \, \sim \,  {\sqrt{\epsilon} \over M_{PL}}
\label{rcondition}
\end{equation}
where $H$ is a Hubble parameter. 
In the approximation of an instant decay (which we adopt for the moment) the reheating 
temperature is given by (\ref{tr}). Since $\lambda_1 > \lambda_2$,
the condition (\ref{rcondition}) gets first satisfied in the 
$\lambda_1$-domain. Thus when
\begin{equation}
 \Gamma_1 = \lambda^2_1m \, \sim \, H 
\end{equation}
the energy in this domain gets converted into
radiation:
\begin{equation}
 \epsilon_{1rad} \,  \sim \,  T_{R1}^4\, \sim \lambda_1^4m^2M_{PL}^2
\end{equation}
which subsequently scales as $1/a^4$.
Right after this moment the energy densities in two domains are still
equal. However, in the $\lambda_1$-domain
it is stored in radiation,
whereas in the $\lambda_2$-domain it is still stored in inflaton oscillations
and evolves as non-relativistic matter. From this moment on
up until  reheating in the $\lambda_2$-domain, the
energy densities in the two domains evolve {\it differently}.

The $\lambda_2$-domain gets reheated when 
\begin{equation}
 \Gamma_2 = \lambda^2_2 \, m \, \sim \, H
\end{equation}
and gets filled with radiation of an energy density 
\begin{equation}
 \epsilon_{2rad} \,  \sim \,  T_{R2}^4\, \sim \lambda_2^4m^2M_{PL}^2
\end{equation}
For equal values of scale factors, the energy in the first domain is redshifted to
\begin{equation}
 \epsilon_{1rad} \,  \sim \, \left ({\lambda_2 \over \lambda_1}\right )^{4/3}
\lambda_2^4m^2M_{PL}^2
\label{e1f}
\end{equation}
The resulting radiation energy densities in these domains are related as
\begin{equation}
 \epsilon_{1rad} \,  \sim  \, \left ({\lambda_2 \over \lambda_1}\right )^{4/3} \epsilon_{2rad}
\label{12}
\end{equation}
The density perturbations are
\begin{equation}
{\delta \epsilon_{rad} \over\epsilon_{rad}} \,  \propto \, 
{ \delta \lambda \over \lambda } \, \propto \, {\delta\, \Gamma \over \Gamma}
\label{deltll}
\end{equation}
Thus, during the reheating process, the fluctuations in $\lambda$
are translated in density perturbations. The precise calculation of the above model (\S 4 and Appendix) actually gives $\delta \epsilon_{rad} / \epsilon _{rad} \approx 0.1\delta \Gamma / \Gamma$.

\section{The origin of the fluctuations in the decay rate}
 
Let us now discuss why the fluctuations in $\Gamma$ 
could be generated in the first place.
The reason why coupling ``constants'' can fluctuate, is that
in unified theories, such as supersymmetric theories, the couplings
are not constant but are determined by expectation values of  fields in the theory.
Many of these fields have very flat potentials and thus can strongly
fluctuate during inflation. 

In general we can think of $\lambda(S)$ as a function of a scalar field
$S$. Let us assume that $q$-s are the ordinary fermions, and their 
supersymmetric partners. The coupling can be
expanded in series of $S$
\begin{equation}
\lambda(S)\, = \, \lambda_0 (1  \, + \, {S \over M} +...)  
\label{expansion}
\end{equation}
Where $M$ is some scale, which may be of order $M_{PL}$ or smaller. 
Since we want to keep our discussion maximally general, we should not specify
the value of $M$.
Depending on the gauge symmetries of the theory the first
(and some of the higher terms) of expansion
may or may not be absent. We shall assume that $S$ is a light field, 
with mass smaller than the Hubble parameter during inflation. Then
$S$ fluctuates during inflation, $\delta S \sim H$  and these 
fluctuations translate
into fluctuations of $\lambda$, which after reheating translate
into the density fluctuations.  

We can define $f$ to be the fraction of the coupling  controlled by the fluctuating field. If the VEV of $S$ during inflation is $\langle S \rangle$ then $f = \langle S \rangle/M$. 
In terms of $f$ we can write
\begin{equation}
{\delta \Gamma \over \Gamma} = f  {\delta S \over \langle S \rangle }
 \end{equation}
The observed level of density perturbation implies that ${\delta \Gamma/  \Gamma} \sim 10^{-5}$. The fluctuations of $S$ are of order, $\delta S \sim H$, so the observed level of fluctuations can be the result of $\langle S \rangle >> H$ or of $f << 1$. 

The fluctuations of $S$ will be Gaussian if $\langle S \rangle >> H$ and non-Gaussian in the opposite limit. Thus depending on the VEV of $S$ during inflation we can get fluctuations with a varying level of non-Gaussianities. Our mechanism for generating non-Gaussianities is different from the multifield inflation models (eg. \cite{finelli, bern} and references therein).

\subsection{Example}

Let us consider a simple illustrative example:
Minimal Supersymmetric Standard Model (MSSM) plus an inflaton $\phi$.
The chiral superfield content of the model consists of two Higgs doublets
(we shall call them $h,\bar{h}$), quark and lepton superfields plus
their anti-particle superfields 
(we shall denote them as $q,q_c$), and an inflaton superfield
($\phi$). Following the usual practice, we shall assume that the inflaton
is a singlet under the MSSM gauge group. 
 The specific form of the inflaton potential is unimportant. 
All we assume is that it gives us a sufficient amount of inflation,
for solving the flatness and horizon problems. We shall also assume
that the density fluctuations
created by the inflaton fluctuations   are very small. 

At the
end of inflation, $\phi$ has to decay and reheat the Universe. 
Let us ask, how could $\phi$ decay into quarks and leptons. 
Since $\phi$ is a gauge singlet, it has three options for 
transmitting its energy into ordinary particles.

{\it 1)} $\phi$ can directly decay into the ordinary particles through the
renormalizable interaction in the superpotential
\begin{equation}
     \lambda_0\, \phi\, h \,\bar{h}
\label{direct}
\end{equation}
In fact due to the gauge symmetry, the only candidates for such a decay
are the Higgs doublets.

{\it 2}) $\phi$ can also decay through non-renormalizable interactions
in the superpotential
\begin{equation}
 \phi \, {q\over M}\, q\, q \, + \phi \, {q_c\over M}\, q_c\, q_c \,
+  \phi \, {h\over M}\, q\, q_c \, +... 
\label{coupling}
\end{equation}
where $M$ is some mass scale. This scale can be $\sim M_{PL}$ or 
may be a lower scale coming 
from integrating out some heavy particles below $M_{PL}$. The precise origin of
the couping is not important.\footnote{Some of the above interactions break 
baryon number, however they cannot lead to proton decay if $\phi$ has 
a zero VEV today. Breaking of baryon number conservation could be even welcome for
generating baryon asymmetry during reheating. This issue will not
be discussed here.} Analogous, couplings may be included in K\"ahler
potential.

{\it 3}) Finally, $\phi$ can decay into some heavy exotic particles
with MSSM gauge charges, which quickly re-scatter and thermalize with the
ordinary particles. In particular, the effective non-renormalizable couplings
of the form (\ref{coupling}) can result from integrating out
such states.

 Let us focus on the second possibility.
If during inflation and reheating one of the sfermions participating in the
the coupling gets a VEV, then through (\ref{coupling}) $\phi$ can 
experience a direct two-body decay into the $q$-quanta. 
The decay rate is regulated by an effective couping
\begin{equation}
 \lambda \, = \, {\langle S \rangle \over M}
\label{hoverm}
\end{equation}
Where  $\langle S \rangle$ is the VEV
of a scalar component of one of the $q$-superfields.
 The resulting partial density perturbations coming from the above
channel will be
\begin{equation}
\delta \epsilon_{through~S} \,  \sim \, 
{ \delta S \over M}\left ({\langle S \rangle \over M}\right )^3m^2M_{PL}^2,
\label{partial1}
\end{equation}
whereas the total density will be sum over all the energies created in
different channels, including the direct decays through non-fluctuating
couplings (\ref{direct})
\begin{equation}
\epsilon_{direct} \,  \sim \, \lambda_0^4m^2M_{PL}^2
\label{partial2}
\end{equation}
(we ignore the delay in thermalization due to re-scattering of the different
MSSM species.)
Thus, the density perturbations will be given by
\begin{equation}
{\delta \epsilon_{rad} \over\epsilon_{rad}} \,  \sim \, 
{\delta \epsilon_{through~S} \over  \epsilon_{through~S} +  \epsilon_{direct}} 
\label{delrho}
\end{equation}

The properties of these perturbations, such as
Gaussianity will be determined by the balance between the different
channels, and on the value of $\delta S$. Let us briefly discuss this issue.

In MSSM
$S$ corresponds to one of the many {\it flat directions} of the potential.
These flat directions are flat only in the unbroken SUSY limit, but are lifted
by SUSY breaking effects, and acquire curvature $\sim$ TeV.

 During inflation (or reheating)
the inflationary energy density breaks supersymmetry
and lifts the flat directions even more, provided $H > $ TeV.
So in general, the flat direction fields acquire masses $\sim H$ and may
acquire big VEVs as well (see \cite{gauge}).

The physical reason behind the $\sim H^2$ curvature
of flat directions during inflation is very transparent. The point is that
a positive vacuum energy density, that drives inflation, breaks supersymmetry
spontaneously. This breaking is either $F$ or a $D$-type, 
meaning that the vacuum
energy density is dominated by either the $F$ or the $D$ terms. In case of,
$F$-breaking, the breaking is universally transmitted to all the fields
through the couplings in the K\"ahler potential, e.g., of the following form 
\begin{equation}
\int \, d^4x \, d^4\theta \, {\phi^*\phi \over M^2} \, S^*S
\label{kahler}
\end{equation}
where $\phi$ is the superfield with a non-zero $F$-term, and
$S$ is a flat direction field. Such couplings will be generated both by
supergravity (in which case $M \sim M_{PL}$) and by gauge 
interactions\cite{gauge} (in which case $M \ll M_{PL}$). For instance,
in the case of a minimal K\"ahler, the gravity-mediated curvature
of the flat directions is
\begin{equation}
m_{gravity}^2 \, = \, 3\, H^2 \, + \, {|W|^2 \over M_{PL}^4}
\label{minimalkahler}
\end{equation}
where $W$ is a superpotential. However,  in the MSSM model the flat directions
are usually lifted also by gauge-mediated contribution. 
For instance, if the inflaton $\phi$ can couple to
some gauge-charged fields (which is usually the case), the
flat directions receive a two-loop gauge contribution
to the masses\cite{gauge}
\begin{equation}
m_{gauge}^2 \, = \, H^2 \,\left ({\alpha \over 4\pi}\right )^2 
\left ({M_{PL} \over |\phi |} \right)^2
\label{gauge}
\end{equation}
One-loop contributions are also possible. The over-all sign of the
contribution and their balance is very model dependent. For instance, in
$D$-term inflation\cite{dterm} gravity-mediated curvature 
is $\ll H^2$, and the gauge-mediated contribution
is dominant. 

 The bottomline of our discussion is that, generically, flat directions are
lifted during inflation by gravitational and gauge effects. As a result
their VEVs are shifted. The location of the minima, and effective 
masses in those minima are
very model dependent and it is impossible to draw a general conclusion.
On the other
hand construction of a particular model which would accommodate our needs
is trivial. The condition under which the new
mechanism of the density perturbations will dominate is that the mass of 
$S$ during inflation be slightly below $H$ (say $\sim 0.1 H$ or so). Then $H$ behaves as practically massless
during inflation and we have $\delta S \sim H$. 

Corrections of the form (\ref{minimalkahler})
would inevitably ruin inflation, and must be avoided. Inflationary scenarios that universally
suppress the gravity-mediated curvature (\ref{minimalkahler})
of flat directions, e.g., such as
the D-term inflation\cite{dterm}, are therefore highly motivated, as they protect
the flatness of the inflaton potential from dangerous 
gravity-mediated corrections 
$\sim H^2$. 
In such scenarios our mechanism will generically be operative and 
contribute to the density perturbations. 

 Then depending on which is the dominant channel of the inflaton decay, we
get different pictures in perturbation spectrum. If couplings like
(\ref{coupling}) are the dominant source of reheating, then the 
density perturbations are given by
\begin{equation}
{\delta \epsilon \over \epsilon} \,  \sim \, 
{ \delta S \over \langle S \rangle }
\label{partial3}
\end{equation}
In which case to get the correct level of density perturbations we have to demand that
$\langle S \rangle \,  \sim 10^{5} \, H$, and resulting perturbations
will be Gaussian, as in the case of an ordinary inflation.

Another possibility is when the channel (\ref{direct}) dominates.
In this case
\begin{equation}
{\delta \epsilon \over \epsilon } \,  \sim \, 
{ \delta S \over M \lambda_0^4} \sim {H \over M \lambda_0^4}
\label{partial4}
\end{equation}
Since now $S$ can stay near its inflationary minimum, the perturbations can be non-Gaussian.

\section{Consequences for model building}

We have argued that our new mechanism for generating density perturbations creates fluctuations of order $H /M$ where $M$ is a scale that could be significantly lower than the Plank scale  (the origin of $M$ depends on the particular model, it could be related to the VEV of a field or with physics at scales intermediate between Hubble during inflation and the Planck Scale).
In other words our mechanism can create the observed density perturbations with a much lower scale of inflation than the standard scenario where density perturbations are given by $H/M_{PL}\sqrt{\epsilon}$, with $\epsilon = (M_{PL}V'/V)^2$.

If our mechanism is responsible for the observed perturbations then we predict a much lower background of gravitational waves than in standard scenarios. The amplitude of the gravity wave background is still given by $H/M_{PL}$ and thus it is greatly suppressed relative to density perturbations, of order $H/M$. 

The relation between the slope of the power spectrum of density perturbations and the characteristics of the inflaton potential is different in our scenario,
\begin{eqnarray}
n-1 &=& {d \ln H^2 \over d \ln a} \ \ \ \ \ \ {\rm our} \ \ {\rm scenario} \nonumber \\ 
n-1 &=& {d \ln H^2/\epsilon \over d \ln a} \ \ \ \  {\rm standard \ \ case}
\end{eqnarray}
 As a result,  both the energy scale of inflation and the characteristics of the inflaton potential would be misinterpreted if our mechanism was operating but the observations were analyzed using the standard assumptions. 

Moreover, there is a significant chance that the fluctuations generated by our mechanism could be  at least slightly non-Gaussian.  The fields that determine the couplings are not slow rolling during inflation and do not dominate the energy density.  As a result, their potentials need not satisfy the stringent slow-roll conditions the inflaton needs to satisfy. These conditions imply that non-Gaussianities generated in standard one-field models of inflation are very small, probably unobservable with current techniques \cite{maldacena}. The fields determining the couplings could have for example larger self couplings 
(such would be, for instance, the quasi-flat directions stabilized by
quartic Yukawa couplings in the superpotential)
leading to observable non-Gaussianities. 

The slow roll condition on the $k$th derivative of the inflaton potential ($V^{(k)}$) is, 
\begin{equation}
\epsilon^{k/2-1} {M_{PL}^k V^{(k)} \over V} < 1.
\end{equation}
For $k=3$ for example, this leads to the  constraint that $V^{(3)}/H < H/M_{PL}\epsilon^{1/2} \sim 10^{-5}$.  In our scenario couplings much larger that this are allowed. The cubic coupling over $H$  is roughly given by $m^2_S / \langle S \rangle H \sim H /\langle S \rangle$ so if  $\langle S \rangle $ is not much larger than $H$ significant non-Gaussianities are possible. 

 The bound on the three point function from WMAP can roughly be translated into $V^{(3)}/H < 10^{-3}$ \cite{wmapnong}.  On the other hand to match the correct level of density perturbations $f \delta S /\langle S \rangle \sim 10^{-4}$, (as we will show in the next section the solution of the full equations imply that the potential fluctuations are of order $1/9 \ \delta \Gamma / \Gamma$ so we set the combination $f \delta S /\langle S \rangle$ to  $10^{-4}$ rather than $10^{-5}$) . This means that when $f \sim 0.1$ the non-Gaussianities  saturate the WMAP bound.  A more detailed analysis of the constraints set by WMAP is left for a future publication. 

\section{Perturbation Equations}

In this section we will explicitly write the relevant perturbation equations in the conformal gauge \cite{conf, mabert}. We will consider two fluids, the inflaton field  which for simplicity we will treat as non-relativistic matter and  radiation.

First we write the equations for the background evolution. The background metric is flat FRW universe, $ds^2=a^2(d\eta ^2-dx^2)$, $\eta$ is the conformal time, $a'/a={\cal H}$ is the conformal Hubble constant, with $'\equiv d/d\eta$. The unperturbed Friedmann equations are

\begin{equation}
\epsilon _m '=-3{\cal H}\epsilon _m-\Gamma a\epsilon _m,
\end{equation}
\begin{equation}
\epsilon _r '=-4{\cal H}\epsilon _r+\Gamma a\epsilon _m,
\end{equation}
\begin{equation}
{\cal H}^2=(8/3)\pi Ga^2(\epsilon _r+\epsilon_m).
\end{equation}

These describe a matter-dominated universe at small time and a radiation-dominated universe at large time. We solved them assuming that  expansion rate was $H_*$ and $x\equiv \epsilon _m/(\epsilon _m+\epsilon _r)=1$ at the beginning. The fraction $x$ remains around one until it suddenly drops to $x=0$ at the time when $H \sim \Gamma$. Before that time $H^2\propto a^{-3}$ and after that   $H^2\propto a^{-4}$. Here $H\equiv{\cal H}/a$ is the standard Hubble parameter.

We now consider {\it super-horizon} perturbations induced by perturbations in the decay rate $\Gamma$. We use the conformal gauge with perturbed metric $ds^2=a^2((1+2\Phi)d\eta ^2-(1-2\Phi)dx^2)$. The perturbed Einstein and matter equations describing forced {\it super-horizon} perturbations are
\begin{equation} \label{forced1}
{\cal H}\Phi '+{\cal H}^2\Phi=-(4/3)\pi Ga^2(\epsilon _m\delta _m+\epsilon _r\delta _r),
\end{equation}
\begin{equation}
\delta _m'=3\Phi'-\Gamma a(\delta _{\Gamma }+\Phi),
\end{equation}
\begin{equation} \label{forced3}
\delta _r'=4\Phi'+(\epsilon _m/\epsilon _r)\Gamma a(\delta _{\Gamma }+\Phi +\delta _m -\delta _r).
\end{equation}
Here $\delta _m\equiv \delta \epsilon _m/\epsilon _m$, $\delta _r\equiv \delta \epsilon _r/\epsilon _r$, $\delta _{\Gamma }\equiv \delta \Gamma /\Gamma$.

\begin{figure}
     \centering
     \centerline{\epsfxsize=8cm\epsffile{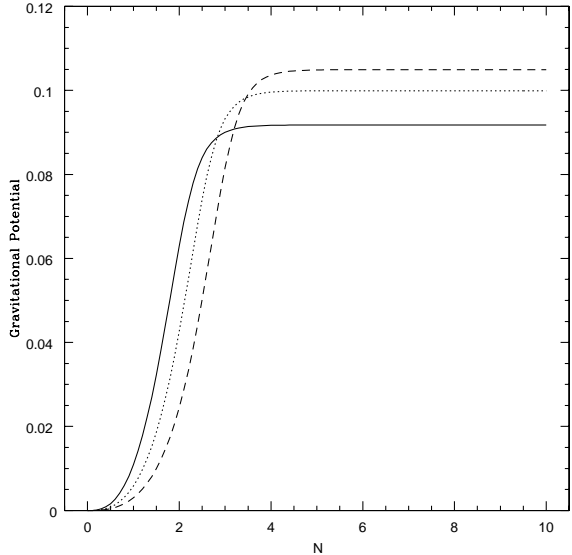}}
     \caption{Evolution of the gravitational potential in units of $\delta \Gamma / \Gamma$ as a function of $N=\ln a$. The curves correspond to  $H_*/ \Gamma = 5, 10, 20$ (solid,dotted,dashed). }
    \label{figure}
\end{figure}

We solved equations (\ref{forced1}-\ref{forced3}) numerically. The solution for the gravitational potential is shown in figure \ref{figure}.  The gravitational potential directly gives the level of Microwave Background anisotropies which are simply proportional to $\Phi$ (see \cite{mabert} for details on the calculation of CMB anisotropies in the conformal gauge).  As we have estimated before, the potential fluctuations are of order $\delta \Gamma /\Gamma$. More specifically $\Phi\approx 1/9 \ \delta \Gamma / \Gamma$  with only  a weak dependance of $ \Gamma/ H_*$, provided $ \Gamma/ H_*$ is relatively small. In the limt $ \Gamma/ H_*\rightarrow 0$, one obtains (Appendix) the exact result $\Phi= (1/9)\delta \Gamma / \Gamma$.

\vspace{0.5cm}   

\section{Conclusions}

We have proposed a new way in which density perturbations could be created during inflation. The density fluctuations produced by the new mechanism could easily dominate over the ones produced in the standard scenario. Our mechanism allows for the energy scale of inflation to be significantly lower than usually assumed. The fluctuations produced are also adiabatic and nearly scale invariant, although the relation of the spectral slope to the inflation potential is different from the usual. If the scale of inflation is lower, we generically predict a lower background of gravitational waves. 

Our mechanism has no trouble generating non-Gaussianities that could be observed by future experiments. If such signal were to be found it would give additional clues as to the details of the inflationary model and how it fits in the general framework of particle physics. 

\begin{flushleft}
{\bf \Large Acknowledgments}\vspace{0.1cm} \\

\end{flushleft}

We thank N. Arkani-Hamed, H.-C. Cheng, P. Creminelli, J. Maldacena, V. F.  Mukhanov, L. Randall, and R. Scoccimarro for useful discussions. 

The work of G.D. is supported in part 
by David and Lucile  
Packard Foundation Fellowship for  Science and Engineering,
by Alfred P. Sloan foundation fellowship and by NSF grant 
PHY-0070787. The work of A.G. is supported in part 
by David and Lucile  
Packard Foundation Fellowship for  Science and Engineering. The work of M. Z. is supported by part by David and Lucile  
Packard Foundation Fellowship for  Science and Engineering.

\begin{appendix}

\section{Appendix: Isomorphism between free and induced perturbations. Exact solution.}

Equations for free perturbations are obtained just by dropping the terms with $\delta _{\Gamma }$ from equations for induced perturbations (\ref{forced1}-\ref{forced3}):
\begin{equation} \label{free1}
{\cal H}\Phi '+{\cal H}^2\Phi=-(4/3)\pi Ga^2(\epsilon _m\delta _m+\epsilon _r\delta _r),
\end{equation}
\begin{equation}\label{free2}
\delta _m'=3\Phi'-\Gamma a\Phi,
\end{equation}
\begin{equation}\label{free3}
\delta _r'=4\Phi'+(\epsilon _m/\epsilon _r)\Gamma a(\Phi +\delta _m -\delta _r).
\end{equation}

Let $\Phi$, $\delta _m$, $\delta _r$ be a solution of the free perturbation equations (\ref{free1}-\ref{free3}). Then $\Phi _1=\Phi -\delta _{\Gamma }$, $\delta _{m1}=\delta _m+2\delta _{\Gamma }$, $\delta _{r1}=\delta _r+2\delta _{\Gamma }$ is a solution of the forced perturbation problem. We are interested in the purely forced mode, when $\Phi _1=\delta _{m1}=\delta _{r1}=0$ at $\eta =0$. The purely forced mode can be obtained from the free solution which satisfies $\Phi =\delta _{\Gamma }$, $\delta _m=-2\delta _{\Gamma }$, $\delta _r=-2\delta _{\Gamma }$ at $\eta =0$. This is just an adiabatic perturbation in the matter-dominated universe. That $\delta _r\neq 0$ at $\eta =0$ is irrelevant, because the radiation energy is small at small $\eta$. 

We will show below that the free adiabatic (not growing for $\eta \rightarrow 0$) perturbation satisfies the well known matching condition \cite{match}
\begin{equation}\label{matching}
{\Phi (\infty )\over \Phi (0)}={1+p_m\over 1+p_r}={10\over 9}.
\end{equation}
Here the indices $p$ characterize the expansion rate of the universe, $a\propto t^p$. For matter domination $p_m=2/3$, and for radiation domination $p_r=1/2$. 

The problem of calculating the forced metric perturbation is thus solved. The fluctuating reaction rate $\delta \Gamma $ induces a super-horizon potential perturbation 
\begin{equation}
\Phi ={1\over 9}{\delta \Gamma \over \Gamma }.
\end{equation}

It remains to prove (\ref{matching}). The free perturbation equations (\ref{free1}-\ref{free3})  can be written as 
\begin{equation}
(3\epsilon _m+4\epsilon _r)^2\zeta '=\epsilon _mS,
\end{equation}
\begin{equation}\label{S}
S'=-{9\epsilon _m+10\epsilon _r\over2\epsilon _m+2\epsilon _r}{\cal H}S,
\end{equation}
where $\zeta$ (the Bardeen parameter) and $S$ are 
\begin{equation}
\zeta =\Phi -{\epsilon _m\delta _m+\epsilon _r\delta _r\over 3\epsilon _m+4\epsilon _r},
\end{equation}
\begin{equation}
S={\cal H}\epsilon_r(3\delta _r-4\delta _m)+\Gamma a(\epsilon _m\delta _m+\epsilon _r\delta _r).
\end{equation}
There are two scalar modes of free perturbations. The mode $S\neq 0$ is obtained from (\ref{S}). The other mode is $S=0$, $\zeta =const$. One can show that the latter mode describes adiabatic perturbations at $\eta \rightarrow 0$. We can therefore use $\zeta =const$ to prove the matching condition (\ref{matching}).

\end{appendix}

\end{document}